\newcommand\ckmfitter{{CKMfitter}}
\newcommand{\simgt}{\,\hbox{\lower0.6ex\hbox{$\sim$}\llap{\raise0.6ex\hbox{$>$}}}\,}
\newcommand{\simlt}{\,\hbox{\lower0.6ex\hbox{$\sim$}\llap{\raise0.6ex\hbox{$<$}}}\,} 
\newcommand{\bbq}{\ensuremath{B_q\!-\!{\bar B}_q\,}}

\newcommand{\ket}[1]{\ensuremath{| #1 \rangle }}

\newcommand{\lt}{\left}
\newcommand{\rt}{\right}
\newcommand{\imag}{\mathrm{Im}\,}

\newcommand{\dm}{\ensuremath{\Delta M}}
\newcommand{\dg}{\ensuremath{\Delta \Gamma}}
\newcommand{\ov}[1]{\overline{#1}}

\newcommand{\braOket}[3]{\langle#1|#2|#3\rangle}
\newcommand{\Bag}{\mathcal{B}}

\documentclass[
  aps,
  prd,
  reprint,
  showpacs,
  groupedaddress,
  amsmath,
  amssymb,
  floatfix
]{revtex4-1}
\usepackage{graphicx,epsfig,dcolumn,multirow}
\newcolumntype{d}[1]{D{.}{.}{#1}}

\begin{document}

\title{
   Predictions of selected flavour observables within the Standard Model }

\author{The \ckmfitter\  Group\\
\vspace{0.1cm}
J.~Charles$^{\,b}$,
O.~Deschamps$^{\,c}$,
S.~Descotes-Genon$^{\,f}$,
R.~Itoh$^{\,e}$,
H.~Lacker$^{\,d}$,
A.~Menzel$^{\,d}$,
S.~Monteil$^{\,c}$,
V.~Niess$^{\,c}$,
J.~Ocariz$^{\,h}$,
J.~Orloff$^{\,c}$,
S.~T'Jampens$^{\,a}$,
V.~Tisserand$^{\,a}$,
K.~Trabelsi$^{\,e}$}

\affiliation{
{ $^{a}$Laboratoire d'Annecy-Le-Vieux de Physique des Particules,\\
                   9 Chemin de Bellevue, BP 110, F-74941
                   Annecy-le-Vieux Cedex, France\\
                   (UMR 5814 du CNRS-IN2P3 associ\'ee \`a
                   l'Universit\'e de Savoie) \\
                {e-mail: tisserav@lapp.in2p3.fr, tjamp@lapp.in2p3.fr}} \\
                { $^{b}$Centre de Physique Th\'eorique, \\
                   Campus de Luminy, Case 907, F-13288 Marseille Cedex 9, France\\
                   (UMR 6207 du CNRS associ\'ee aux
                   Universit\'es d'Aix-Marseille I et II \\et
                   Universit\'e du Sud Toulon-Var; laboratoire
                   affili\'e \`a la FRUMAM-FR2291) \\
                {e-mail: charles@cpt.univ-mrs.fr}} \\
                { $^{c}$Laboratoire de Physique Corpusculaire de Clermont-Ferrand \\
                  Universit\'e Blaise Pascal,
                  24 Avenue des Landais F-63177 Aubiere Cedex \\
                  (UMR 6533 du CNRS-IN2P3 associ\'ee \`a
                   l'Universit\'e Blaise Pascal) \\
                  {e-mail: odescham@in2p3.fr, orloff@in2p3.fr, monteil@in2p3.fr, niess@in2p3.fr}} \\
                  { $^{d}$Humboldt-Universit\"at zu Berlin,\\
                   Institut f\"ur Physik,
                   Newtonstr. 15,\\ 
                   D-12489 Berlin, Germany \\
                {e-mail: lacker@physik.hu-berlin.de, amenzel@physik.hu-berlin.de}}\\
                { $^{e}$High Energy Accelerator Research Organization, KEK \\
                  1-1 Oho, Tsukuba, Ibaraki 305-0801 Japan \\
                {e-mail: ryosuke.itoh@kek.jp, karim.trabelsi@kek.jp}}\\
                { $^{f}$Laboratoire de Physique Th\'eorique \\
                   B\^{a}timent 210, Universit\'e  Paris-Sud 11, F-91405 Orsay Cedex, France \\
                   (UMR 8627 du CNRS  associ\'ee \`a l'Universit\'e Paris-Sud 11) \\
                {e-mail: Sebastien.Descotes-Genon@th.u-psud.fr}}\\{ $^{h}$Laboratoire de Physique Nucl\'{e}aire et de Hautes Energies, \\
IN2P3/CNRS, Universit\'{e} Pierre et Marie Curie Paris 6 \\ et Universit\'{e} Denis
Diderot Paris 7, F-75252 Paris, France \\
{e-mail: Ocariz@in2p3.fr}}}

\date{\today}

\begin{abstract}
This letter gathers a selection of Standard Model predictions issued from the metrology of the CKM parameters performed by the CKMfitter group. The selection includes purely leptonic decays of neutral and charged $B$, $D$ and $K$ mesons. In the light of the expected measurements from the LHCb experiment, a special attention is given to the radiative decay modes of $B$ mesons as well as to the $B$-meson mixing observables, in particular the semileptonic charge asymmetries $a^{d,s}_{\rm SL}$ which have been recently investigated by the {D\O} experiment at Tevatron.          
Constraints arising from rare kaon decays are addressed, in light of both current results and expected performances of future rare kaon experiments.
All results have been obtained with the \ckmfitter\ analysis package, featuring the frequentist statistical approach and using Rfit to handle theoretical uncertainties.
\end{abstract}

\pacs{12.15.Hh,12.15.Ji, 12.60.Fr,13.20.-v,13.38.Dg}

\maketitle

\begin{table*}
\renewcommand\arraystretch{1.2}
\caption{Constraints used for the global fit, and the main inputs involved (more information can be found in ref.~\cite{CKMfitterwebsite}). The lattice inputs are our own averages obtained as described in the text. \label{tab:expinputs}}
\begin{tabular}{c|c|cccc|ccc}
CKM  & Process  & \multicolumn{4}{c|}{Observables}  & \multicolumn{3}{c}{Theoretical inputs}\\
\hline
$|V_{ud}|$ & $0^+\to 0^+$ transitions 
                  & $|V_{ud}|_{\rm nucl}$&=& $0.97425\pm 0.00022$
                  & \cite{Hardy:2008gy} & \multicolumn{3}{c}{Nuclear matrix elements} \\
                    \hline
$|V_{us}|$ & $K\to\pi\ell\nu$ 
                  & $|V_{us}|_{\rm semi}f_+(0)$&=& $0.2163 \pm 0.0005$ & \cite{FlaviaNet}
                  & $f_+(0)$&=& $0.9632\pm 0.0028\pm 0.0051$\\
                 &  $K\to e\nu_e$ 
                 & ${\cal B}(K\to e\nu_e)$&=&$ (1.584\pm 0.0020)\cdot 10^{-5}$ & \cite{PDG}
                 &  $f_K$&=& $156.3\pm 0.3\pm 1.9$ MeV \\
                &  $K\to \mu\nu_\mu$ 
                &  ${\cal B}(K\to \mu\nu_\mu)$&=& $0.6347\pm 0.0018$
                & \cite{FlaviaNet}\\
                 &  $\tau \to K \nu_\tau$ 
                 & ${\cal B}(\tau \to K\nu_\tau)$&=&$0.00696\pm 0.00023$
                 & \cite{PDG}\\
                 \hline
$|V_{us}|/|V_{ud}|$                 &  $K\to \mu\nu/\pi\to\mu\nu$ & 
                 $\displaystyle \frac{{\cal B}(K\to \mu\nu_\mu)}{{\cal B}(\pi \to \mu\nu_\mu)}$
                         &=&$(1.3344\pm 0.0041)\cdot 10^{-2}$
                 & \cite{FlaviaNet} &
                 $f_K/f_\pi$&=&$1.205\pm 0.001 \pm 0.010$ 
                  \\
                 &  $\tau\to K\nu/\tau \to \pi\nu$ &   
                 $\displaystyle \frac{{\cal B}(\tau \to K\nu_\tau)}{{\cal B}(\tau \to \pi\nu_\tau)}$
                        &=& $(6.53\pm 0.11)\cdot 10^{-2}$
                 & \cite{Banerjee:2008hg} \\
                 \hline
$|V_{cd}|$ & $D\to \mu\nu $ & ${\cal B}(D\to \mu\nu)$ &=& $(3.82\pm 0.32\pm 0.09)\cdot 10^{-4}$ 
                   & \cite{CLEO:2008sq}
                   &$f_{D_s}/f_D$&=&$1.186\pm 0.005 \pm 0.010$\\
                   \hline
$|V_{cs}|$   & $D_s\to \tau\nu$ 
                   & ${\cal B}(D_s\to \tau\nu)$&=& $(5.29\pm 0.28)\cdot 10^{-2}$ 
                   &  \cite{Asner:2010qj} 
                   & $f_{D_s}$ &=& $251.3 \pm 1.2 \pm 4.5$ MeV\\
                   & $D_s\to \mu\nu$ 
                   & ${\cal B}(D_s\to \mu\nu_\mu)$&=& $(5.90\pm 0.33)\cdot 10^{-3}$ 
                   &  \cite{Asner:2010qj}\\
                   \hline
$|V_{ub}|$ & semileptonic decays 
                  & $|V_{ub}|_{\rm semi}$ &=& $(3.92\pm 0.09\pm 0.45)\cdot 10^{-3}$ 
                  &  \cite{Asner:2010qj}&
                     \multicolumn{3}{c}{form factors, shape functions}\\
                  & $B\to \tau\nu$ 
                  & ${\cal B}(B\to\tau\nu)$ &=& $(1.68\pm 0.31)\cdot 10^{-4}$ & \cite{CKMfitterwebsite}
                  & $f_{B_s}$ &=& $ 231\pm 3\pm 15$ MeV\\
                  &&&&&&  $f_{B_s}/f_B$&=& $1.209\pm 0.007\pm 0.023$\\
                  \hline
$|V_{cb}|$ & semileptonic decays
                 & $|V_{cb}|_{\rm semi}$ &=& $(40.89\pm 0.38 \pm 0.59)\cdot 10^{-3}$ &  \cite{Asner:2010qj}
                 &  \multicolumn{3}{c}{form factors, OPE matrix elements}\\
\hline
$\alpha$ & $B\to\pi\pi$, $\rho\pi$, $\rho\rho$ 
                & \multicolumn{3}{c}{branching ratios, CP asymmetries} & \cite{Asner:2010qj} 
                & \multicolumn{3}{c}{isospin symmetry}\\
                \hline
$\beta$   & $B\to (c\bar{c}) K$ 
               & $\sin(2\beta)_{[c\bar{c}]}$ &=& $0.678\pm 0.020$ 
               & \cite{Asner:2010qj}\\
\hline
$\gamma$ & $B\to D^{(*)} K^{(*)}$ 
                 & \multicolumn{3}{c}{inputs for the 3 methods}
                 &  \cite{Asner:2010qj}& \multicolumn{3}{c}{GGSZ, GLW, ADS methods} \\
 \hline
$V_{tq}^*V_{tq'}$       & $\Delta m_d$ 
                & $\Delta m_d$ &=& $0.507\pm 0.005$ ps${}^{-1}$
                & \cite{Asner:2010qj}
                &  $\hat{B}_{B_s}/\hat{B}_{B_d}$ &=& $1.01\pm 0.01\pm 0.03$\\ 
                 & $\Delta m_s$ & $\Delta m_s$ &=& $17.77\pm 0.12$ ps${}^{-1}$ 
                 & \cite{Abulencia:2006ze} 
                 & $\hat{B}_{B_s}$&=& $1.28\pm 0.02\pm 0.03$\\
                 \hline
$V_{tq}^*V_{tq'},V_{cq}^*V_{cq'}$  
      & $\epsilon_K$ & $|\epsilon_K|$ &=& $(2.229\pm 0.010)\cdot 10^{-3}$
       & \cite{PDG} 
       &$\hat{B}_K$&=& $0.730\pm 0.004\pm 0.036$\\
    &   &&&&& $\kappa_\epsilon$&=& $0.940\pm 0.013\pm 0.023$
\end{tabular}

\end{table*}

\begin{figure}
\includegraphics[width=9cm]{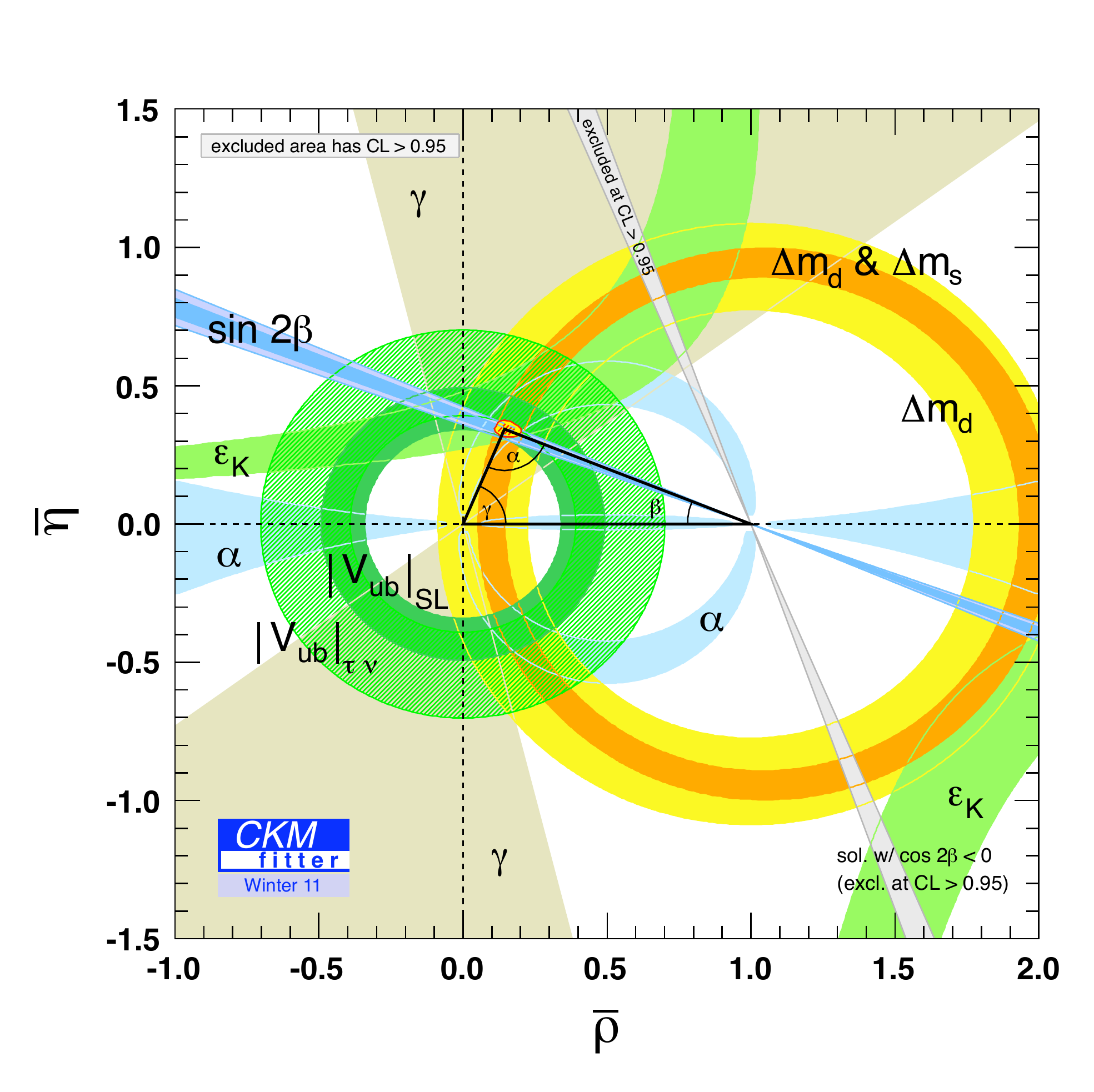}
\caption{Constraints on the CKM $(\bar\rho,\bar\eta)$ coordinates from the global SM
CKM-fit. Regions outside the coloured areas have $CL > 95.45~\%$. For the combined
fit the yellow area inscribed by the contour line represents points with $CL < 95.45~\%$.
The shaded area inside this region represents points with $CL < 68.3~\%$.\label{fig:global fit}}
\end{figure}

\section{Introduction}

In the Standard Model (SM), the weak charged-current transitions
mix quarks of different generations, which is encoded in the
unitary Cabibbo-Kobayashi-Maskawa (CKM) matrix~\cite{Cabibbo:1963yz,Kobayashi:1973fv}.
In the case of three generations of quarks,
the physical content of this matrix reduces to four real parameters,
among which one phase, the only source of $CP$ violation in the SM
(we do not consider minute $CP$-violating effects from the strong-interaction $\theta$-term or from the masses of neutrinos). 
One can define these four real parameters as:
\begin{eqnarray}
\lambda^2&=&\frac{|V_{us}|^2}{|V_{ud}|^2+|V_{us}|^2}\,,
\quad
A^2\lambda^4 =\frac{|V_{cb}|^2}{|V_{ud}|^2+|V_{us}|^2}\,, \nonumber\\
\bar\rho+i\bar\eta&=&-\frac{V_{ud}V_{ub}^*}{V_{cd}V_{cb}^*}\,,
\end{eqnarray}
and exploit the unitarity of the CKM matrix to determine all its elements (and when needed, to obtain their expansion in powers of $\lambda$)~\cite{Charles:2004jd}.

Extracting information on these parameters from data is a challenge for both experimentalists and theorists, since
the SM depends on a large set of parameters which are not predicted within its framework, and must be determined experimentally. A further problem comes from the presence of the
strong interaction that binds quarks into hadrons and is still difficult to tackle theoretically, leading to most of the theoretical uncertainties discussed when extracting the CKM matrix parameters. The CKMfitter group follows this goal using a standard $\chi^2$-like frequentist approach, in addition to the Rfit scheme to treat theoretical uncertainties, aiming at combining a large set of constraints from flavour physics~\cite{Charles:2004jd,CKMfitterwebsite}.

Not all the observables in flavour physics can be used as inputs for these constraints, due to limitations on our experimental and/or theoretical knowledge on these quantities. The list of inputs of the global fit is indicated in table~\ref{tab:expinputs}: they fulfill the double requirement of a satisfying control of the attached theoretical uncertainties and a good experimental accuracy of their measurements. In addition, we only take as inputs the quantities that provide reasonably tight constraints on the CKM parameters $A,\lambda,\bar\rho,\bar\eta$.
This selects in particular leptonic and semileptonic decays of mesons yielding information on moduli of CKM matrix elements, non-leptonic two-body $B$ decays related to angles of the CKM matrix, and $B$ and $K$-mixing parameters.

The current situation of the global fit in the $(\bar\rho,\bar\eta)$ is indicated in Fig.~\ref{fig:global fit}. Some comments are in order before discussing the metrology of the parameters. There exists a unique preferred region defined by the entire set of observables under consideration in the global fit. This region is represented by the yellow surface inscribed by the red contour line for which the values of $\bar\rho$ and $\bar \eta$ correspond to $CL < 95.45~\%$. The goodness of the fit can be addressed in the simplified case where all the inputs uncertainties are taken as Gaussian, with a $p$-value found to be $14 \%$ (i.e., 1.5 $\sigma$; a rigorous derivation of the $p$-value in the general case is beyond the scope of this letter~\cite{wip}).  One obtains the following values (at 1 $\sigma$) for the 4 parameters describing the CKM matrix: 
\begin{eqnarray}
A= 0.816^{+0.011}_{-0.022}\,, &\qquad&
\lambda=0.22518^{+0.00036}_{-0.00077}\,\\
\bar\rho = 0.144^{\,+0.028}_{\,-0.019}\,, &\qquad&
\bar\eta = 0.342^{\,+0.015}_{\,-0.014}\,.
\end{eqnarray}

At this stage, it is fair to say that the SM hypothesis has passed the statistical test of the global consistency of all observables embodied in the fit, although some discrepancies are detailed in the following sections.  We are therefore allowed to perform the metrology of the CKM parameters and to give predictions for any CKM-related observable within the SM.  Let us add that the existence of a  $CL < 95.45~\%$ region in the $({\bar\rho,\bar \eta})$ plane is not equivalent to the statement that each individual constraint lies in the global range of $CL < 95.45~\%$.  One of the interest of SM predictions is that each comparison between the prediction issued from the fit and the corresponding measurement constitutes a null-test of the SM hypothesis. Indeed, we will see that discrepancies actually do exist among  the present set of observables considered in this letter (the corresponding pulls are reported in Table~\ref{tab:pred:meas}).

We predict observables that were not used as input constraints, either because they are not measured with a sufficient accuracy yet, e.g., ${\cal B}(B_s\to\ell^+\ell^-)$, or because the control on the theoretical uncertainties remains controversial, e.g., $\Delta\Gamma_s/\Gamma_s$. The corresponding predictions can then be directly compared with their experimental measurements (when they are available). We also consider some particularly interesting observables used as an input of the fit, e.g., ${\cal B}(B\to\tau \nu)$. In this last case, we must compare the measurement of the observable with the outcome of the fit without including the observable among the inputs, so that the experimental information is used only once.

Following this procedure, we do not take the following quantities as inputs, but we predict their values:
the semileptonic asymmetries $a^s_{\rm SL}$ and $a^d_{\rm SL}$, the weak phase in the $B^0_s$ mixing $\beta_s$, the branching ratios  of the dileptonic decays of neutral $B$ mesons ${\cal B} (B_{d,s} \to \ell^+ \ell^- $), the branching ratio of (exclusive and inclusive) radiative $b\to s$ transitions, and rare $K\to\pi\nu\bar\nu$ decays. The first three observables have all in common to provide only loose constraints on the CKM parameters, while the two latter, though fulfilling the  requirement of a good control of their related  theoretical uncertainties, are so far out of reach of the current experiments.  The LHCb experiment should bring a breakthrough in that respect very soon and these quantities will be included in the global fit once the required measurement accuracy is achieved~\cite{LHCb:2009ny}.  The experimental situation is pretty similar for the semileptonic asymmetries related to neutral-meson mixing, with the additional drawback that these observables suffer from large theoretical uncertainties. The exclusive radiative $b \to s$ transitions suffer from significant uncertainties and are thus only consider for predictions. On the contrary, the inclusive $B\to X_s\gamma$, which have been measured and are well controlled theoretically, will be added as input of the global fit~\cite{wip}, but are kept for the present letter among the predictions. Finally, rare kaon decays have not been measured yet or provide only loose constraints on the CKM matrix elements.

In the following sections, we first discuss the main sources of theoretical uncertainty, before spelling out some of the fundamental formulae used for our predictions within the SM. We then collect the results obtained and compare them with their measurements (when available).

\section{Strong interaction parameters}

The first category of theoretical uncertainties in flavour analyses arise from matrix elements that encode the effects of strong interaction in the non-perturbative regime.
These matrix elements boil down to decay constants, form factors and bag parameters for most of the observables under scrutiny in the present note, and all our predictions are subjected
to and limited by the uncertainties in the determination of these observables. These uncertainties must be controlled with care since their misassessment or underestimation would
affect the statements that we will make on flavour observables.

Among the different methods used to estimate non-perturbative QCD parameters, quark models, sum rules, and lattice QCD (LQCD) simulations are tools of choice. We opt for the latter whenever possible, as they provide well-established methods to compute these observables not only with a good accuracy at the present time, but also with a theoretical framework allowing for a systematic improvement on the theoretical control of the uncertainties. 
Over the last few years, many new estimates of the decay constants and the bag factors have been issued by different lattice collaborations, with different ways to address the uncertainties. A part of the uncertainties has a clear statistical interpretation: lattice simulations evaluate Green functions in an Euclidean metric expressed as path integrals using Monte Carlo methods, whose accuracy depends on the size of the sample of gauge configurations used for the computation in a straightforward way. But systematics are also present and depend on the strategies of computation chosen by competing lattice collaborations: discretisation methods used to describe gauge fields and fermions on a lattice, interpolating fields, parameters of the simulations, such as the size of the (finite) volumes and lattice spacing, the masses of the quarks that can be simulated, and the number of dynamical flavours included as sea quarks (2 and 2+1 being the most frequent, after a long period where only quenched simulations were available). These simulations must be extrapolated to obtain physical quantities (relying in particular on effective theories such as chiral perturbation theory and heavy-quark effective theory).

The combination of lattice values with different approaches to address the uncertainty budget is a critical point of most global analyses of the flavour physics data, even though the concept of the theoretical uncertainty for such quantities is ill-defined (and hence is the combination of them). Several approaches have been proposed to perform such a combination.
We have collected the relevant LQCD calculations of the decay constants $f_{B_d}$, $f_{B_s}$, $f_{D_s}$, $f_{D}$, $f_{K}$, $f_{\pi}$, as well as the bag parameters ${\cal B}_{B_d}$, ${\cal B}_{B_s}$ and ${\cal B}_{K}$, and the $K_{\ell 3}$ form factor $f_+(0)$. In addition we designed an averaging method aiming at providing a systematic, reproducible and to some extent conservative scheme~\cite{Tisserand:2009ja}. These lattice averages are the input parameters used in the fits presented in this paper.

In the specific case of decay constants, the $SU(3)$-flavour breaking ratios 
$f_K/f_\pi$, $f_{D_s}/f_D$, $f_{B_s}/f_{B_d}$ are better determined than the individual decay
constants. We will therefore take these ratios as well as the strange-meson decay constants
as reference quantities for our inputs. In the same spirit,
it is more relevant to consider the predictions of the ratio $K_{\ell 2}/\pi_{\ell 2}$ of the kaon and pion leptonic partial widths, as well as $\mathcal{B}(\tau\to K\nu_\tau)/\mathcal{B}(\tau\to \pi\nu_\tau)$ instead of the individual branching ratios. 

A comment is in order concerning the second category of theoretical uncertainties that are not directly related to LQCD. As far as global fit inputs are concerned, this is the case for  the inclusive and exclusive determinations of $|V_{ub}|$ and $|V_{cb}|$, which involve non-perturbative inputs of different nature. We use the latest HFAG results~\cite{Asner:2010qj} for each of these determinations and combine inclusive and exclusive determinations following the same scheme as for the combination of lattice quantities. We refer the reader to refs.~\cite{
Lenz:2010gu,Deschamps:2009rh} for a more detailed discussion of each constraint, whereas the related hadronic inputs can be found in ref.~\cite{CKMfitterwebsite}.

\section{Neutral $\mathbf{B}$-meson leptonic decays}

Dileptonic decays of $B_d$ and $B_s$ mesons are among the most appealing
laboratories to study scalar couplings in addition to the SM
couplings. The current experimental limits set by the Tevatron and LHCb
experiments on the dimuonic branching ratios~\cite{TeVexp,Aaij:2011rj} are 
already giving significant constraints on scenarios beyond the Standard 
Model. The main limitation in the current predictions arises from the
knowledge of the decay constants $f_{B_d}$ and $f_{B_s}$.
\\
The master formula for the branching ratios reads as: 
\begin{equation}
\begin{split}
&{\cal B}[ B_{d,s} \rightarrow \ell^+ \ell^-]_{\rm SM}  = \frac{G_F^2 \alpha_{em}^2  f_{B_{d,s}}^2 m_{\ell}^2  m_{B_{d,s}} \tau_{B_{s,d}} }{ 16 \pi^3 \sin^4\theta^{\rm eff}_W } \\
 &\qquad\qquad \times \sqrt{ 1 - \frac{4m_{\ell}^2}{m_{B_{d,s}}^2} } |V^*_{tb}V_{t(d,s)}|^2\ Y^2\left(\frac{\ov m_t^2}{m_W^2}\right),
\end{split}
\label{eq_btoll}
\end{equation}	
	 where $Y$ is the next-to-leading-order Inami-Lim function~\cite{BBL,bsmumuSM}. $\ov m_t$ is the top quark mass defined in the $\ov{\text{MS}}$ scheme,
related to the pole mass $m_t^\text{pole}$ determined at the Tevatron as
$\ov m_t(\ov m_t) =  0.957 m_t^\text{pole}$  at next-to-leading order
of QCD. $G_F$ is  the Fermi constant, $\sin^2\theta^{\rm eff}_W$ the electroweak mixing angle, and $\alpha_{em}$ the electromagnetic coupling constant at the $Z$ pole. We vary the renormalisation scale between $m_t/2$ and $2m_t$.

\section{Charged meson leptonic decays}

The decay of a charged meson $M$ ($=\pi,K,D,D_s,B$) into a leptonic pair $\ell \nu_\ell$ is mediated in the SM by a charged weak boson, with the branching ratio:  

%
\begin{equation}
\begin{split}
{\cal B}[M\rightarrow \ell \nu_\ell]_{\rm SM} & = \frac{G_F^2 m_M m_{\ell}^2 }{ 8 \pi }  \left( 1 - \frac{m_{\ell}^2}{m_M^2} \right)^2 |V_{q_uq_d}|^2 \\
                                              &  \quad\quad \times f_M^2 \tau_M ( 1 + \delta_{em}^{M \ell 2} ),
\end{split}
\end{equation}	
where $q_u$ ($q_d$) stands for the up-like (down-like) valence quark
of the meson respectively, $V_{q_uq_d}$ is the relevant CKM matrix
element, $f_M$ is the decay constant of the meson $M$ and
$\tau_M$ its lifetime. A similar formula holds for $\tau$ decays into a single light meson (pion or kaon).
 The corrective factor $\delta_{EM}^{M \ell 2}$
stands for channel-dependent electromagnetic radiative
corrections. They are taken into account in the case of the lighter
mesons ($\pi$ and $K$), where their impact is estimated to be at the
level of $2-3\%$~\cite{Kl2pil2em}, and for the $D$ mesons, where the effect is~$1\%$~\cite{Dl2em}. 
In the case of $B$ mesons, no dedicated corrections
are supposed here, and we assume that all the corrections from soft photons will be taken into account
through the Monte Carlo analyses of the experiments (see ref.~\cite{Bl2em} for a discussion of the corrections due to soft-photon emission).

\section{Radiative $\mathbf{B}$-meson decays}

The radiative transitions $b\to s(d) \gamma(^*)$ provide very interesting probes 
of New Physics, as they are mediated by penguin transitions which are directly related
in the SM with $B_d$ and $B_s$ mixing (from the CKM point of view), but can
be affected differently by additional particles/couplings. A convenient
framework for their analysis is provided by the effective Hamiltonian where
all degrees of freedom heavier than the $b$-quark have been integrated out, leading to Wilson coefficients $C_i$ (encoding short-distance physics) multiplied by operators with light degrees of freedom (describing long-distance physics).

Hadronic uncertainties may be significant for the exclusive decays ${\cal B} (B_d \to K^*(892) \gamma)$ and ${\cal B} (B_s \to \phi \gamma)$ due to the form factors and the long-distance gluon exchanges:
\begin{equation}
\begin{split}
&{\cal B}[\bar{B}\to V\gamma]_{\rm SM}
 = \frac{\tau_B}{c_V^2}\frac{G_F^2\alpha_{em} m_B^3m_b^2}{32\pi^4}
 \left(1-\frac{m_V^2}{m_B^2}\right)^3 \\ & \qquad
   \times  [T_1^{B\to V}(0)]^2 \sum_{X=L,R} \left|\sum_{U=u,c} V_{Us}^* V_{Ub} a_{7X}^U\right|^2\,,
\end{split}
\label{Eq:BVgamma}
\end{equation}
where $c_V$ is a Clebsch-Gordan coefficient, $\alpha_{em}$ the electromagnetic coupling constant at vanishing momentum, the index $X=L,R$ corresponds to the photon polarisation, and
$a_{7R}^{c,u}$ is $m_s/m_b$ suppressed compared to $a_{7L}^{c,u}=C_7+O(\alpha_s,1/m_b)$ where $C_7$ is the Wilson coefficient of the electromagnetic dipole operator (corrections can be estimated using a $1/m_b$ expansion). In eq.~(\ref{Eq:BVgamma}), $U$ denotes any up-type quark. We follow ref.~\cite{Ball:2006eu} for the prediction of the branching ratio and the analysis of hadronic uncertainties (however, we use results from light-cone QCD sum rules and do not perform any rescaling of the tensor form factor $T_1^{B\to V}(0)$).

The inclusive transition $b\to s\gamma$ can be treated relying on the quark-hadron duality and using a heavy-quark expansion, so that the prediction for this transition suffers from fewer theoretical uncertainties (mostly related to the precise values of the quark masses and the higher orders in both $\alpha_s$ and $1/m_b$ expansions). However, this observable is not fully inclusive as a cut in the photon energy is required. The corresponding expression is~\cite{bsgamma}:
\begin{equation}
\begin{split}
&{\cal B}[\bar{B} \to X_s\gamma]_{{\rm SM},E_\gamma > E_0}
 ={\cal B}[\bar{B}\to X_c e\bar\nu]\\ 
 &\qquad\qquad \times
    \left|\frac{V_{ts}^* V_{tb}}{V_{cb}}\right|^2
   \frac{6\alpha_{em}}{\pi C} [P(E_0)+N(E_0)]\,,
   \end{split}
\end{equation}
where $C$ is a factor related to the choice of $b\to c$ transition as a normalisation for the computation, $N(E_0)$ collects the estimate from
non-perturbative $1/m_b$-suppressed contributions, and $P(E_0)$ has been estimated up to next-to-next-to-leading order using an interpolation on the charm quark mass, leading
to a formula of the form
$P(E_0)=\sum_{i,j} C_i C_j K_{ij}(E_0)$ where the $K_{ij}$ are perturbative kernels. For the present analysis, we use the parametrisation described in detail in ref.~\cite{Deschamps:2009rh}.

The (exclusive and inclusive) radiative decays $b\to s\ell^+\ell^-$ provide  more observables, which are
already experimentally accessible, but they are out of the scope of this short note~\cite{wip}.

\section{$\mathbf{CP}$-violating $\mathbf{B}$-mixing observables}

The mixing of the $B_d$ and $B_s$ mesons can be described upon introducing 
the mass and decay matrices, $M^q=M^{q\dagger}$ and
$\Gamma^q=\Gamma^{q\dagger}$. These matrices are involved in the evolution
operator for the quantum-mechanical description of the  \bbq\
oscillations (with $q=d$ or $q=s$).
Their diagonalisation defines the physical eigenstates $\ket{B^q_H}$ and
$\ket{B^q_L}$ with masses
$M^q_H,\,M^q_L$ and decay rates $\Gamma^q_H,\,\Gamma^q_L$.
One can reexpress these quantities in terms of three parameters:   $|M_{12}^q|$,
$|\Gamma_{12}^q|$ and the relative phase $\phi_q=\arg(-M_{12}^q/\Gamma_{12}^q)$.

Oscillation frequencies, which feature the CKM elements directly,
can be predicted in a theoretically clean way, though the precision is severely limited by
the knowledge of the decay constants and bag parameters~\cite{Lenz:2010gu}. Here we would
like to concentrate on the prediction of the four
$CP$-violating observables $\beta_q$ (mixing phases)
and  $a^q_{\rm SL}$ (semileptonic asymmetries), with $q=d$ or $q=s$.  The two
first observables are CKM angles of the $B_d$ and $B_s$  unitarity
triangles, respectively, and read as functions of the CKM elements:
\begin{eqnarray}
\beta &=& \arg \left (-\frac{V_{cd}V_{cb}^*}{V_{td}V_{tb}^*} \right ),\\
\beta_s &=& -\arg \left (-\frac{V_{cs}V_{cb}^*}{V_{ts}V_{tb}^*}\right ).
\end{eqnarray}
These angles (which should not be confused with the relative phases $\phi_q$ introduced above) 
measure $CP$ violation arising in the
interference between  mixing and decay in $b\to c\bar{c}s$ and hence exhibit a
strong hierarchy between the $d$ and $s$ quarks.
On the contrary, the semileptonic asymmetry probes $CP$
violation in the mixing, and can be written as:
\begin{eqnarray}\nonumber
a^q_{\rm SL} &=&  2\, \lt( 1- \lt| \frac{q}{p}\rt| \rt) \; =\; \imag \frac{\Gamma_{12}^q}{M_{12}^q} \\
&= & \frac{|\Gamma_{12}^q|}{|M_{12}^q|} \sin \phi_q \; = \; \frac{\dg_q}{\dm_q} \tan \phi_q
. \label{defafs}
\end{eqnarray}
The ingredients needed to predict these asymmetries are hence the
matrix elements $M_{12}$ and $\Gamma_{12}$. The dispersive term
$M_{12}^q$ is mainly driven by box diagrams involving virtual top
quarks, and it is related to the effective $|\Delta B|=2$ Hamiltonian
$H_q^{|\Delta B|=2}$ as:
\begin{eqnarray}\label{eq:m12q}
M_{12}^q &=& \frac{\braOket{B_q}{H_q^{|\Delta B|=2}}{\ov B_q}}{2 M_{B_q}}
\, .
\end{eqnarray}
The SM expression for $H_q^{|\Delta B|=2}$ is \cite{BBL}:
\begin{eqnarray}
H_q^{|\Delta B|=2} &=& (V_{tq}^* V_{tb}^{\phantom{*}})^2 \, C \, Q \; +\;
\mbox{h.c.}\label{defheff}
\end{eqnarray}
with the four-quark operator
$Q = \ov q_L \gamma_\mu b_L \, \ov q_L \gamma^\mu b_L$ and the Wilson coefficient $C$:
\begin{eqnarray}
C^{\rm SM} &=& \frac{G_\text{F}^2}{4\pi^2} M_W^2 \, \widehat \eta_B\,
S\lt( \frac{\ov m_t^2}{M_W^2} \rt) . \label{defc}
\end{eqnarray}
 and the Inami-Lim function $S$ is calculated from the box
diagram with two internal top quarks. 
The absorptive term $\Gamma_{12}^q$ is dominated by
on-shell charmed intermediate states, and it can be expressed as a
two-point correlator of the $|\Delta B|=1$ Hamiltonian $H_q^{|\Delta
B|=1}$. By performing a $1/m_b$-expansion of this two-point
correlator, one can express $\Gamma_{12}^q$ in terms of $Q$ and 
$\widetilde{Q}_S  =   \ov{q}_L^\alpha b_R^\beta \,
\ov{q}_L^\beta  b_R^\alpha$,
where $S$ stands for ``scalar'' and $\alpha,\beta=1,2,3$ are colour indices~\cite{Lenz:2010gu}.  The matrix elements are expressed in terms of the bag parameters:
\begin{eqnarray}
\braOket{B_q}Q{\ov B_q} &=& \frac{2}{3} M^2_{B_q}\,
f^2_{B_q} \widetilde \Bag_{B_q},\\
\braOket{B_q}{\widetilde{Q}_S}{\ov B_q} &=& 
\frac{1}{12}  M^2_{B_q}\,
f^2_{B_q} \widetilde \Bag_{S,B_q}^\prime\,.
\label{defbstp}
\end{eqnarray}
One has also to consider further  bag parameters $\Bag_{R,B_q}$ 
 which parameterise the
matrix elements  of the subleading operators in the heavy quark
expansion of the $CP$-violating observables (only rough estimates are available for these bag
parameters)~\cite{Lenz:2010gu}. The SM predictions of the mixing phases and semileptonic asymmetries for the neutral mesons in the SM are collected in table~\ref{tab:pred:meas}.

\begin{table*}

\renewcommand\arraystretch{1.2}

\caption{Comparison between prediction and measurement of some flavour observables
in the SM. The first
  column describes the observables. The second and third columns give the measurement and the
  prediction from the global fit (not including the measurement of the quantity considered),
  respectively. The fourth column expresses the departure of the
  prediction to the measurement, when available. 
   \label{tab:pred:meas}}
  \begin{tabular}{c|cc|rl|c}
   Observable & \multicolumn{2}{c|}{Measurement} & \multicolumn{2}{c|}{Prediction}  & Pull ($\sigma$)
     \rule[-2mm]{0pt}{4ex} \\ 
  
   \hline
   \multicolumn{6}{c}{Charged Leptonic Decays \rule[-2mm]{0pt}{4ex} } \\
   \hline 
   ${\cal B}(B^+\to \tau^+\nu_{\tau}) $ & $ (16.8 \pm 3.1) \cdot 10^{-5}$  & \cite{CKMfitterwebsite} 
   &  (7.57 & ${}^{+0.98}_{-0.61})\cdot 10^{-5}$   & 2.8 \\ 
   ${\cal B}(B^+\to \mu^+\nu_{\mu})$	& $< 10^{-6}$   & \cite{Asner:2010qj} 
   &  (3.74  & ${}^{+0.44}_{-0.38})\cdot 10^{-7}$   & - \\ 
   ${\cal B}(D_s^+\to \tau^+\nu_{\tau}) $	& $(5.29 \pm 0.28)\cdot 10^{-2}$ &  \cite{Asner:2010qj} 
   &  (5.44  & ${}^{+0.05}_{-0.17})\cdot 10^{-2}$   & 0.5 \\ 
   ${\cal B}(D_s^+\to \mu^+\nu_{\mu}) $	& $(5.90  \pm 0.33)\cdot 10^{-3}$ &    \cite{Asner:2010qj} 
   &  (5.39  & ${}^{+0.21}_{-0.22})\cdot 10^{-3}$   & 1.3 \\ 
   ${\cal B}(D^+\to \mu^+\nu_{\mu}) $	& $(3.82  \pm 0.32  \pm 0.09)\cdot 10^{-4}$ &   \cite{CLEO:2008sq}
   & (4.18  & ${}^{+0.13}_{-0.20})\cdot 10^{-4}$   & 0.6 \\ 
   \hline
   \multicolumn{6}{c}{Neutral Leptonic $B$ decays  \rule[-2mm]{0pt}{4ex}} \\
   \hline
  ${\cal B}(B^0_s \to \tau^+\tau^-)$	&  -  &
   &  (7.73  & ${}^{+0.37}_{-0.65})\cdot 10^{-7}$   &  -\\ 
   ${\cal B}(B^0_s \to \mu^+\mu^-)$	& $<   32\cdot 10^{-9}$  & \cite{Asner:2010qj}
   &  (3.64  & ${}^{+0.17}_{-0.31})\cdot 10^{-9}$   &  -\\ 
  ${\cal B}(B^0_s \to e^+ e^-)$	&  $< 2.8\cdot 10^{-7}$    &  \cite{Asner:2010qj} 
   &  (8.54  & ${}^{+0.40}_{-0.72})\cdot 10^{-14}$   & - \\
   ${\cal B}(B^0_d \to \tau^+\tau^-)$	& $<4.1  \cdot 10^{-3}$  & \cite{Asner:2010qj} 
   &  (2.36  & ${}^{+0.12}_{-0.21})\cdot 10^{-8}$   & - \\ 
  ${\cal B}(B^0_d \to \mu^+\mu^-)$	& $< 6\cdot 10^{-9}$  & \cite{Asner:2010qj} 
   &  (1.13  & ${}^{+0.06}_{-0.11})\cdot 10^{-10}$   & -\\ 
   ${\cal B}(B^0_d \to e^+ e^-)$	& $<$ $8.3\cdot 10^{-9}$  &  \cite{Asner:2010qj} 
   &  (2.64  & ${}^{+0.13}_{-0.24})\cdot 10^{-15}$   &  -\\ 
 
   \hline
   \multicolumn{6}{c}{$\bbq$ mixing observables \rule[-2mm]{0pt}{4ex}} \\
   \hline
     $\Delta\Gamma_s/\Gamma_s$ & $0.092^{+0.051}_{-0.054}$  & \cite{Asner:2010qj} 
   & 0.179  & ${}^{+0.067}_{-0.071}$   & 0.5 \\
  $a_{\rm SL}^d  $	& $(-47 \pm 46)\cdot 10^{-4}$  & \cite{Asner:2010qj} 
     & (\,-6.5 & ${}^{+1.9}_{-1.7}\ \,)\cdot 10^{-4}$   & 0.8 \\ 
   $a_{\rm SL}^s $	& $(-17 \pm 91  {}^{+12}_{-23})\cdot 10^{-4}$  &\cite{Abazov:2009wg}
   & (0.29  & ${}^{+0.09}_{-0.08})\cdot 10^{-4}$   & 0.2 \\ 
 $a_{\rm SL}^s-a_{\rm SL}^d$	& -  &
   & (\ 6.8  & ${}^{+1.9}_{-1.7}\ \,) \cdot 10^{-4}$   & - \\ 
 $\sin(2\beta)$ & 0.678 $\pm$ 0.020 &   \cite{Asner:2010qj} 
   &  0.832 & ${}^{+0.013}_{-0.033}$   & 2.7 \\
\multirow{2}{*}{$2\beta_s$} & $[0.04; 1.04]\cup  [2.16; 3.10]$ & \cite{CDFbetas}
   &  \multirow{2}{*}{0.0363} & \multirow{2}{*}{${}^{+0.0016}_{-0.0015}$}   & \multirow{2}{*}{-} \\
   & 0.76  ${}^{+0.36}_{-0.38}$  $\pm$ 0.02 & \cite{D0betas} & & & \\
   \hline
   \multicolumn{6}{c}{Radiative $B$ decays \rule[-2mm]{0pt}{4ex}} \\
   \hline
    ${\cal B}(B_d\to K^*(892) \gamma)  $	& $(43.3 \pm 1.8)\cdot 10^{-6}$  & \cite{Asner:2010qj} 
   & (64 & ${}^{+22}_{-21})\cdot 10^{-6}$  & 1.2 \\
    ${\cal B}(B^-\to K^{*-}(892) \gamma)$	& $(42.1 \pm 1.5)\cdot 10^{-6}$   & \cite{Asner:2010qj}
   & (66  & ${}^{+21}_{-20})\cdot 10^{-6}$   & 1.1 \\
     ${\cal B}(B_s\to  \phi \gamma) $ & $(57^{+21}_{-18})\cdot 10^{-6}$  & \cite{Asner:2010qj} 
   &  (65  & ${}^{+31}_{-24})\cdot 10^{-6}$   & 0.1 \\
  ${\cal B}(B\to X_s \gamma) $/	${\cal B}(B\to X_c\ell\nu)$& $(3.346 \pm 0.247)\cdot 10^{-3}$ &   \cite{Asner:2010qj}
   &  (3.03  & ${}^{+0.34}_{-0.32})\cdot 10^{-3}$   & 0.2 \\
     \hline
   \multicolumn{6}{c}{Rare $K$ decays \rule[-2mm]{0pt}{4ex}} \\
   \hline
    ${\cal B}(K^+ \to \pi^+\nu\bar\nu)$	& $(1.75^{+1.15}_{-1.05})\cdot 10^{-10}$   & \cite{E949}
   &  (0.854  & ${}^{+0.116}_{-0.098})\cdot 10^{-10}$   & 0.8 \\
  ${\cal B}(K_L \to \pi^0\nu\bar\nu) $	&  - & 
   &  (0.277  & ${}^{+0.028}_{-0.035})\cdot 10^{-10}$  & - \\
 \end{tabular}
\end{table*}

\section{The rare kaon decays $K^+\rightarrow\pi^+\nu\bar{\nu}$ and $K_L\rightarrow\pi^0\nu\bar{\nu}$}

Theoretically clean constraints on the CKM matrix can be obtained from
rare kaon decays  with neutrinos in the final state, as they can only arise
via second-order weak transitions ($Z$-penguins and box) within the SM, and
light-quark loops are strongly GIM-suppressed.
Within the SM, the $K^+\rightarrow\pi^+\nu\bar{\nu}$ decay rate is given by~\cite{Buchalla:1998ba,Buras:2005gr,Brod:2008ss,Isidori:2005xm}:
\begin{eqnarray}
&&{\cal  B}[K^+\to\pi^+\nu\bar{\nu}]_{\rm SM} = \kappa_+\left(1+\Delta_{em}\right)
\left[\left(\frac{Im\lambda_t}{\lambda^5}X_t\right)^2\right.\nonumber\\
&&\qquad \qquad
\left.+\left(\frac{Re\lambda_c}{\lambda}\left(P_c+\delta P_{c,u}\right)+\frac{Re\lambda_t}{\lambda^5}X_t\right)^2\right],
\end{eqnarray}
where the isospin-breaking parameter $\kappa_+$ can be extracted from semileptonic $K$ decays with a correction $\Delta_{em}$ for the photon cut-off dependence,
the $X_t$ functions comprise the top quark contributions, 
and the light quark contributions are given by the $P_c$ and $\delta P_{c,u}$ parameters, which are the dominant theoretical uncertainties.
Similarly for the $K_L\rightarrow\pi^0\nu\bar{\nu}$ mode, the SM decay rate is given by~\cite{Buchalla:1998ba}:
\begin{eqnarray}
{\cal  B}[K_L\to\pi^0\nu\bar{\nu}]_{\rm SM} = \kappa_L \left(\frac{Im\lambda_t}{\lambda^5}X_t\right)^2,
\end{eqnarray}
with only  small residual uncertainties from the isospin-breaking parameter $\kappa_L$ and scale invariance. 
In terms of CKM parameters, a measurement of the $K^+\rightarrow\pi^+\nu\bar{\nu}$ provides a quasi-elliptical constraint in the $(\bar{\rho},\bar{\eta})$ plane,
centered close to the vertex of the unitarity triangle located at $(1,0)$. The measurement of the branching ratio for $K_L\rightarrow\pi^0\nu\bar{\nu}$ 
would provide a clean constraint on $\bar{\eta}^2$.

\section{Discussion}

This letter collects a selection of SM predictions
driven by the global fit of the CKM parameters, in view of related
recent or foreseeable experimental measurements.   The main outcome is summarised
in Table~\ref{tab:pred:meas}, gathering the SM predictions using the inputs collected
in Table~\ref{tab:expinputs}. The third column of Table~\ref{tab:pred:meas} shows the
agreement between the measurement and the prediction as a pull. The latter is computed
from the $\chi^2$ difference with and without the measurement of this observable, interpreted with the appropriate
number of degrees of freedom, and converted in the number of equivalent standard
deviations (the lack of an updated average for $\beta_s$ between the Tevatron experiments explain the presence of two distinct measurements as well as the absence of a pull).
 
The largest departures of the measurements from the SM predictions are found for two
 observables: ${\cal B}(B^+ \to \tau^+ \nu_{\tau})$ and $\sin(2\beta)$. It is remarkable that
 this discrepancy can be accommodated by a very simple extension of the SM allowing
 for the  presence of New Physics in $B$ mixing, as discussed extensively in ref.~\cite{Lenz:2010gu}. One can also 
 notice that the $D_s\to\mu\nu$ decay exhibits only a mild discrepancy between prediction and measurement, due to the recent improvements in both lattice simulations and experimental measurements.

Concerning neutral-meson mixing, and following the outstanding success of the $B$ factories  in their
measurements of $\sin (2 \beta)$, one of the main goals and challenges for
for the LHCb experiment will consist in characterising
 the $B$-meson mixing properties through the measurement of
all the relevant observables. Each of these measurements in LHCb but $\sin (2 \beta)$ 
will provide a null test of the SM hypothesis. 
In the present experimental context, two of these observables are particularly interesting: the $B^0_s \overline {B}^0_s$ mixing weak phase $\beta_s$ and the difference of the semileptonic asymmetries for the $B_d$ and $B_s$ mesons. The former is predicted very accurately:     
\begin{eqnarray}
 2\beta_s = 0.0363 {}^{+0.0016}_{-0.0015}\,.
\end{eqnarray}
Significant constraints have already been set on this phase by the Tevatron experiments~\cite{CDFbetas,D0betas}. The LHCb experiment should in a near future settle its value, as suggested by the promising exploratory work with the first data described in ref.~\cite{LHCbbetas}.  

The semileptonic asymmetries are determined far less precisely  by the global fit of the CKM parameters. Their prediction suffers from notable strong-interaction uncertainties (in particular bag parameters). Yet, following a recent D{\O} measurement of the dimuon asymmetry which departs from the SM by 3.2 $\sigma$~\cite{Abazov:2010hv}, the measurement by the LHCb experiment of the difference of the semileptonic asymmetries $a_{\rm SL}^s-a_{\rm SL}^d$ is eagerly awaited. The prediction of the difference in the SM is:     
\begin{eqnarray}
a_{\rm SL}^s-a_{\rm SL}^d = (6.8^{+1.9}_{-1.7}) \cdot 10^{-4}\,.  
\end{eqnarray}
Among the null tests of the SM hypothesis, the $Z$-penguin decay rate ${\cal B} (B^0_s \to \mu^+ \mu^-)$ is specially appealing. Its next-to-leading order prediction from the global fit reads:
\begin{equation}
{\cal B}(B^0_s \to \mu^+\mu^-) = (3.64^{+0.17}_{-0.31})\cdot 10^{-9}\,.
\end{equation}

We would like to conclude this discussion with observables which can uniquely be measured at super-$B$ factories. The important role of ${\cal B}(B^+\to \tau^+\nu_{\tau})$ onto the global fit has been already underlined in this letter, and its SM prediction is: 
\begin{equation}
{\cal B}(B^+\to \tau^+\nu_{\tau}) =  (7.57^{+0.98}_{-0.61})\cdot 10^{-5}\,.
\end{equation}
 An improved precision of the measurement can only be achieved at high-luminosity $B$ factories. The branching ratio of the muonic mode, predicted to be:
 \begin{equation}
 {\cal B}(B^+\to \mu^+\nu_{\mu})=(3.74 {}^{+0.44}_{-0.38})\cdot 10^{-7}\,,
 \end{equation}
 is a further experimental target. 

Let us finally add that this short letter has collected the SM predictions for some salient observables in flavour physics, in view of the running or foreseen experimental programmes here. This obviously does not exhaust the discussion of the inputs, predictions and methods dealt with the  \ckmfitter\  package, but
we leave this subject for a more extensive forthcoming publication~\cite{wip}.

\end{document}